\RequirePackage{fix-cm}
\documentclass[twocolumn]{svjour3}          
\smartqed  
\usepackage{pgfplots,amsfonts,amsmath,amssymb,dcolumn,bm,fleqn,multirow,subfigure,microtype}
\usepackage[%
  bookmarksnumbered=true,
  colorlinks=true,%
  breaklinks=true,%
  citecolor=black,%
  allcolors=black,%
  linkcolor=black,%
  urlcolor=black,
  pdfauthor={Dennis Hessling}
  ]{hyperref}%
\usepackage{siunitx}
\usetikzlibrary{patterns,calc,shadows.blur,shadings,external}
\tikzset{external/system call={lualatex \tikzexternalcheckshellescape --enable  -write18 --shell-escape -tok_size=200000000  -jobname "\image" "\texsource"}}%
  \SendSettingsToPgf
\pgfdeclarelayer{background}
  \pgfdeclarelayer{foreground}
  \pgfsetlayers{background,main,foreground}
\usepackage[utf8x]{inputenc}
\definecolor{M2ired}      {rgb}{0.8000,0.0100,0.2700}
  \definecolor{M2igray}     {rgb}{0.8000,0.8000,0.8000}
  \definecolor{M2iblue}     {rgb}{0.5608,0.7098,0.8000}
\usepackage{lipsum}
\pgfplotsset{compat=newest}
\def\new#1{#1}

\newcommand{\force}{\vec{F}}
\newcommand{\mass}{m}
\newcommand{\Pos}{\vec{x}}

\newcommand{\distf}{f} 
\newcommand{\ls}{\Delta x}
\newcommand{\rt}{\tau} 
\newcommand{\sos}{c_s}
\newcommand{\ts}{\Delta t}
\newcommand{\Lbvv}{\vec{c}} 
\newcommand{\LBw}{w} 
\newcommand{\LBu}{\vec{u}} 
\newcommand{\sccc}{g} 

\begin{document}
\title{Coalescence of immersed droplets on a substrate}
\author{Dennis Hessling \and Jacco H. Snoeijer \and Jens Harting}
\institute{Dennis Hessling \at Materials innovations institute (M2i), Elektronicaweg 25, 2628 XG Delft,Netherlands\\
Department of Applied Physics, Eindhoven University of Technology, 
Postbus 513, 5600 MB Eindhoven, The Netherlands
\and
Jacco H. Snoeijer \at Department of Applied Physics, Eindhoven University of Technology, 
Postbus 513, 5600 MB Eindhoven, The Netherlands\\
    Faculty of Science and Technology, University of Twente, 
Drienerlolaan 5, 7522 NB Enschede, The Netherlands
\and
Jens Harting \at Forschungszentrum J\"ulich GmbH, Helmholtz Institute 
    Erlangen-N\"urnberg for Renewable Energy (IEK-11), F\"urther Stra{\ss}e 248, 
90429 N\"urnberg, Germany
\email{j.harting@fz-juelich.de}\\
    Department of Applied Physics, Eindhoven University of Technology, 
Postbus 513, 5600 MB Eindhoven, The Netherlands\\
    Faculty of Science and Technology, University of Twente, 
Drienerlolaan 5, 7522 NB Enschede, The Netherlands
}
\maketitle

\begin{abstract}
    \new{The dynamics of droplets on substrates has a strong impact on
microfluidic systems ranging from commercially available lab-on-chip systems to state of the art developments in open microfluidics.}
    Coalescence of micro and nano droplets on a substrate has been studied extensively\new{, but in previous studies the focus has been on the interface movement.}
Here, we use computer
simulations to investigate coalescence of droplets immersed in another liquid,
in an inertia-dominated regime \new{and also investigate the droplet's internal flow field}. It is found that qualitatively the dynamics is
similar to coalescence in air, with 
\new{the same}
self-similar growth laws.
We here point out the
ambiguity in the scaling argument for droplets of \SI{90}{\degree}, that shows
itself in the velocities.  We show that droplets with a contact angle below
\SI{90}{\degree} exhibit a self-similar velocity field, and the corresponding
scaling laws are identified.  For drops of \SI{90}{\degree}, however, it is
shown that the velocity field has a more intricate structure that is beyond the
usual scaling arguments invoked for coalescence.
\PACS{%
47.55.df, 
47.55.D-, 
47.85.mb 
}
\end{abstract}

\maketitle

\section{Introduction}
\new{Commercially available microfluidic devices generally consist of closed
channels. These are of advantage in order to prevent evaporation and
allow to pump fluids by applying a pressure difference between inflow and
outflow boundaries. However, keeping the channels clean is a serious problem
due to clogging. Open microfluidic systems are an alternative route, where
droplets and rivulets are confined to chemically patterned substrates
containing wetting surface domains on a less wetting
substrate~\cite{DOERFLER,DOERFLER-24,DOERFLER-25,DOERFLER-26,DOERFLER-27,DOERFLER-28}. While flow in chemical channels cannot be induced by pressure differences, capillary forces due to wettability gradients~\cite{DOERFLER-29}, or electrowetting~\cite{DOERFLER-33} are a possible alternative. Another practical possibility are shear forces induced by a covering immiscible fluid~\cite{DOERFLER-41}. The advantage of this approach is also that it prevents evaporation and contamination, e.g. with dirt or dust.}
\new{In microfluidic systems and in open systems in particular,  
the separation and coalescence of micro and nano droplets is a common phenomenon. However, it occurs not just in microfluidics, but in various
situations, such as cloud formation, film formation~\cite{TenNij00}
or inkjet printing~\cite{KarRie13,LiuMeiWan14,ThoTipJue14}. } 
    While sessile droplets with a contact angle of \SI{90}{\degree} are
    comparable to freely suspended droplets, sessile droplets of different
    contact angles are omnipresent and can have a significantly altered
    behaviour, as we show in this work.
Many different
phenomena are found in this process, like jumping of the coalescing
droplets~\cite{LiuGhiFen14}, a transition from coalescence to
noncoalescence~\cite{KarHanFerRie14,KarRie13}, variations in the meniscus
shape~\cite{BilKin05} and resulting behavior~\cite{HerLubEdd12},
mixing~\cite{SunMac95,ZhaObeSwa15} or the growth rate dependence of the
meniscus on the contact angle~\cite{SuiMagSpe13,GroSteRaa13}.  Furthermore, a
large interest in droplets and bubbles submerged in a liquid
exists~\cite{LohZha15,BerPosGun84,BhuWanMaa08,BorDamSch07}.  Often experiments
on droplets are performed submerged~\cite{BaiWhi88}, in order to reduce the
influence of gravity, to scale diffusion~\cite{SuNee13,DucNee06} or in order to
study altered friction behavior~\cite{Raa04,SchHar11}. 

\begin{figure}[h]\centering
\includegraphics{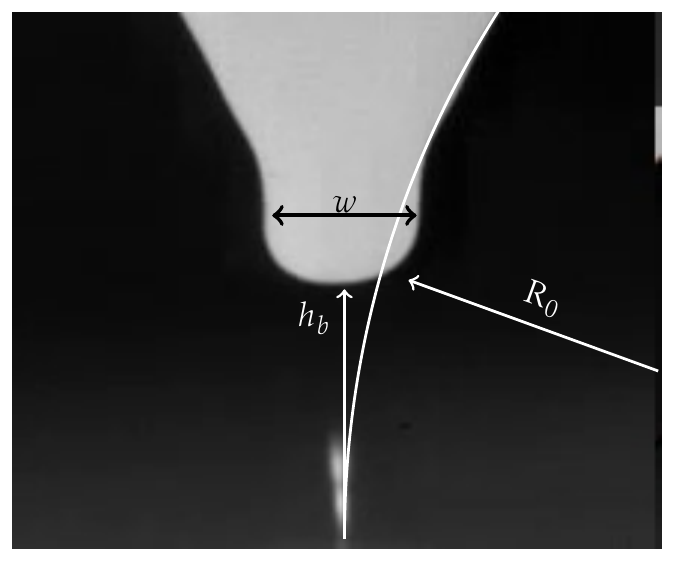}
\caption{Snapshot of the bridge shape during coalescence from the experiment by
    Eddi~et~al.~\cite{EddWinSno13-2}. In this example the droplets have a
    contact angle of \SI{90}{\degree}. The bridge height $h_b$, initial droplet
    radius $r_0$ and the horizontal scale $w$ are marked in the image.
    \label{fig:Experiment}} \end{figure}

Of particular interest is the initial coalescence dynamics, just after the drops
are brought into contact. The two drops become connected by a small liquid
bridge that rapidly grows in time (Fig.~\ref{fig:Experiment}). This rapid motion
is due to the large curvature that induces a very large (negative) Laplace
pressure, driving the liquid into the bridge between the two drops. This process
has for example been studied extensively for freely suspended drops
\cite{PauBurNag11,PauBurNag211,EggLisSto99,DucEggJos03,AarLekGuo05,SprShi14}.
Here we concentrate on drops on a substrate, which further complicates the
geometry of the coalescence process
\cite{RisCalRoy06,NarBeyPom08,LeeKanYoo12,HerLubEdd12,EddWinSno13-2,SuiMagSpe13,MitMit15}.
An important observation is that the dominant direction of the flow is oriented
from the center of the drops towards the bridge, such that the relevant scaling
laws can be inferred from quasi-two-dimensional arguments
\cite{RisCalRoy06,NarBeyPom08,HerLubEdd12,EddWinSno13-2,SuiMagSpe13}.
Interestingly, for inertia-dominated coalescence it was shown that droplets of a
\SI{90}{\degree} contact angle behave differently from those of a lower contact
angle. Even small deviations from \SI{90}{\degree} lead to a faster growth of
the bridge height in time, which can be described in terms of scaling laws. In
particular, the inertial coalescence changes from $t^\frac{1}{2}$ for
\SI{90}{\degree}, to a $t^\frac{2}{3}$ power law for smaller angles
\cite{EddWinSno13-2,SuiMagSpe13}.  In both cases the shape of the liquid bridge
exhibits a self-similar dynamics, but with different horizontal and vertical
scales when the contact angle reaches \SI{90}{\degree}. 

In this paper we focus on the coalescence of sessile drops immersed in another
liquid, as it is relevant in open microfluidics. This has been addressed in the
viscous regime \cite{MitMit15}, for which both the drops and the surrounding
fluid were highly viscous. 
    Here we perform lattice Boltzmann simulations in the inertial regime and
    define the outer fluid to be of the same density as the coalescing
    droplets.  
We investigate how the bridge dynamics changes as a function of the
contact angle and, to the best of our knowledge, for the first time investigate
whether self-similar behavior can be identified in the velocity field. A
detailed comparison to experiments~\cite{EddWinSno13-2} of drops in air is
provided, pointing out similarities and differences with respect to immersed
droplets. The paper starts with a  description of the lattice Boltzmann method,
and we pay particular attention to the initiation of the coalescence
(Sec.~\ref{sec:method}). The central results are presented in
Sec.~\ref{sec:results} and the paper closes with a discussion in
Sec~\ref{sec:discussion}.

\section{Simulation Method}\label{sec:method}

\subsection{The lattice Boltzmann method}

The coalescence of droplets is a quasi 2D problem 
\cite{RisCalRoy06,HerLubEdd12,EddWinSno13-2}, so in favor of numerical speed 
we choose to perform our simulations in 2D. 
To investigate the coalescence of droplets on a substrate~\cite{DOERFLER}, we use the lattice 
Boltzmann method (LBM)~\cite{Raa04,Suc01,BenSucSau92} in a D2Q9 
configuration~\cite{Sri14,QiaYueSuc95}, that can be described by 
\begin{multline}
   f_{i}^\alpha(\Pos+\Lbvv_i \ts,t+\ts) - \distf_{i}^\alpha(\Pos,t)=\\
   -\frac{1}{\rt^\alpha} \left(\distf_i^\alpha(\Pos,t) - 
   \distf_{i_{eq}}^\alpha(\Pos,t)\right),\label{eq:lbm}
\end{multline}
where $\distf_i^\alpha(\Pos,t)$ is a probability distribution function of
particles with mass $\mass$ of component $\alpha$ at position $\Pos$ and time
$t$, following the discretized velocity direction $\Lbvv_i$.  The left hand
side of \eqref{eq:lbm} is the streaming step, where the probability
distribution functions of the fluid $\alpha$ is distributed to the surrounding
lattice sites. The timestep $\ts$, the lattice constant $\ls$ and the mass
$\mass$ of this process are chosen unity for simplicity.  On the right hand
side of \eqref{eq:lbm}, the collision step, these distributions relax towards
an equilibrium distribution
\begin{multline}
    \distf_{i_{eq}}^\alpha(\Pos,t)=\\ \LBw_i \rho^\alpha \left[ 
        1+\frac{\Lbvv_i\cdot \LBu}{\sos^2} + 
        \frac{(\Lbvv_i\cdot\LBu^\alpha)^2}{\sos^2}
        -\frac{(LBu^{\alpha})^2}{2\sos^2}\right]
\end{multline}
on a timescale determined by the relaxation time $\rt$.  The relaxation
time is directly proportional to the kinematic viscosity as
$\nu^\alpha=\frac{2\rt^\alpha-1}{6}$. For simplicity $\rt^\alpha$ is chosen unity here.
Forces can
be added by shifting the equilibrium distribution function and thereby
implicitly adding an acceleration~\cite{ShaChe94}.  Multiple components may
coexist on every lattice site.  Via forces, these can interact with each other.
Here we follow the method described by Shan and Chen~\cite{ShaChe94}
\begin{equation}
   \force^\alpha (x,t) =
    -  \rho^\alpha(\Pos,t) \sccc^{\alpha\overline{\alpha}} \sum\limits_{i=1}^{9}  
    \rho^{\overline{\alpha}}(\Pos+c_i,t)c_i.\label{eq:SCForcePsifun1}
\end{equation}
These interaction forces cause the separation of fluids and a surface-tension
$\gamma$. Here we restrict ourselves to two fluids and refer to them as ``red''
and ``blue'' fluids. The width of fluid interfaces and the resulting surface
tension are governed by the interaction strength parameter
$\sccc^{\alpha\overline{\alpha}}$, which is chosen as $0.9$ for all shown
simulations. This results in a surface tension of $\gamma=1.18 \frac{\ls
\mass}{ {\ts}^{2}}$. 
Our choice of parameters implies that the viscous length scale 
$\frac{\rho^\alpha{\nu^\alpha}^2}{\gamma}$
is comparable to the lattice unit $\Delta x$. The resulting scale for the
coalescence is thus much larger than the viscous length, ensuring we are in the
inertial regime of coalescence~\cite{EggLisSto99,PauBurNag11,EddWinSno13-2}.

In this setup the droplets sit on a horizontal flat substrate.  The horizontal
no slip boundary sites $w$ defining the substrate are modified to include a
pseudo-density~\cite{SchHar11} equal to the average of the surrounding fluid
sites. Interactions as described in \eqref{eq:SCForcePsifun1}, with interaction
parameters $\sccc^{\alpha w}$ and $\sccc^{\overline{\alpha} w}$ cause a contact
angle~\cite{HuaHauTho07,SchHar11}.  In equation \eqref{eq:SCForcePsifun1}
$g^{\alpha w}$ and $g^{\overline{\alpha} w}$ act in place of the interaction
strength parameter and scale the absolute interaction force of the wall on the
fluids, but $\sccc^{\alpha w} - \sccc^{\overline{\alpha} w}$ defines the
contact angle~\cite{HuaHauTho07}.  The parameters are chosen as $g^{\alpha w} =
- g^{\overline{\alpha} w}$ to minimize absolute forces.

Using full slip boundaries to effectively mirror the system at the symmetry 
axis, the computational domain and therefore computational cost is halved.
The computational costs have to be considered for the two following reasons:
as we are interested in obtaining the meniscus height over time with sufficient 
accuracy, we need to scale our entire system to a large resolution.
A second numerical effect to consider is that, because of the finite width of the 
interface, the interface and the resulting fluid behavior can be 
overrepresented.  To avoid this, the interface thickness should be small, as 
compared to the droplet radius~\cite{GroSteRaa13}.
The droplet radius at a contact angle of \SI{90}{\degree} is chosen to be $900 
\ls$, so that the interface thickness results in about \SI{0.5}{\percent} of the 
droplet radius. It was found empirically that this drop size is sufficient to 
achieve reproducible results for the resulting hydrodynamics.
All fluid volumes and system dimensions are kept constant and only the wall 
interaction parameters $\sccc^{\alpha w} = - g^{\overline{\alpha} w}$ and a 
horizontal shift length, that brings the droplets into contact, are adjusted for 
all subsequent simulations of decreasing contact angles.

\subsection{Initialization of coalescence}

Both in experiment and in simulations, it is challenging to initiate coalescence and 
to define the time $t=0$ that marks the start of the coalescence. 
Here we provide technical details on how the simulations were performed. 

A first problem is that the width of the diffuse interface and the fluid
pressures are dependent on the $\sccc^{\alpha\overline{\alpha}}$ fluid
parameters, so that it is not possible to predict beforehand the correct
densities at all positions. This is why equilibration during initialization is
required.  
A lack of correct initialization will lead to strong artefacts, such as
an enclosed bubble for droplets 
of a \SI{90}{\degree} contact angle that can be avoided with careful initialization. 
To
do so, we first equilibrate a horizontally centered single droplet at the
wetting wall before a second drop is introduced.

\begin{figure}
    \subfigure{
      \label{fig:simsketch}
  \includegraphics[width=\dimexpr0.95\linewidth+4\subfigcapmargin]{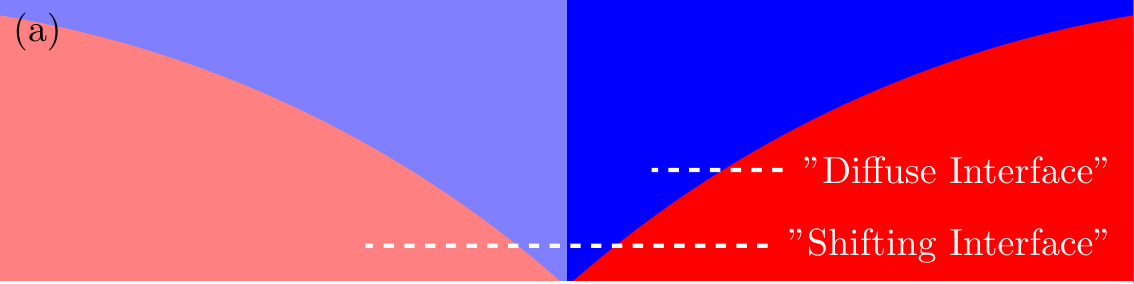}
  }\\
   \subfigure{
      \label{fig:gradl}
  \includegraphics[width=.45\linewidth]{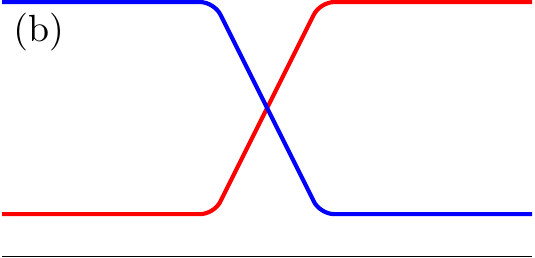}
   }
   \subfigure{
      \label{fig:gradtwo}
  \includegraphics[width=.45\linewidth]{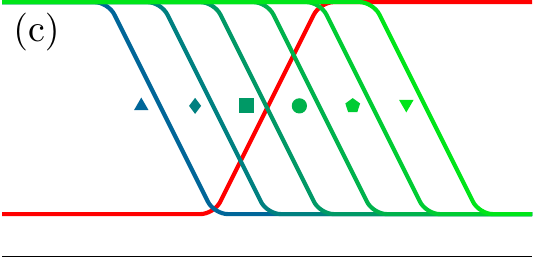}
   }
\caption{(Color online) 
\subref{fig:simsketch} Schematic drawing of droplets on a substrate, before 
coalescence. The opaque half is mirrored through a free slip boundary. The 
density line cross section of Fig.~\subref{fig:gradl} is indicated by the dashed line ``Diffuse 
Interface'' and the corresponding cross section of \subref{fig:gradtwo} is 
indicated with ``Shifting Interfaces''.
\subref{fig:gradl} Schematic of the density of red and blue fluid in a 1D 
cross section across the diffuse interface, as can be found along the dashed line ``Diffuse Interface'' in \subref{fig:simsketch}.
The width of the diffuse interface is about six lattice sites.
\subref{fig:gradtwo} The density cross section at the interface after moving the 
droplet to initialize coalescence. Only the density of the droplets is shown. 
Different possible options to shift the droplet are depicted by a left-right 
transition of the right droplet's density field. The symbols correlate to those 
in Fig.~\ref{fig:tNot}. 
}
\end{figure}

The introduction of a second drop, and the initiation of coalescence is a subtle
matter by itself.  Here we shift the droplet to a system boundary with a full
slip boundary condition.  This effectively mirrors the droplet, as is depicted
in the schematic figure~\ref{fig:simsketch}. This is a magnified section of the
system at the meniscus after shifting the droplet.  Here, the effectively
mirrored part is shown in slightly opaque and density profiles at different
cross sections of the diffuse interface are sketched in Figs.~\ref{fig:gradl}
and \ref{fig:gradtwo}.  The mirrored part of the system shown in
Fig.~\ref{fig:simsketch} does not need to be simulated, which reduces the
simulation time. The density gradients shown in Fig.~\ref{fig:gradl} of the two
fluids exemplary show a transition from a majority of one fluid to the other, as
for instance in the 1D cross section in Fig.~\ref{fig:simsketch}, marked with
"Diffuse Interface".  This schematic representation of a diffuse interface
identifies that the position of an according sharp interface is not clearly
defined.  Accordingly, the shift to the full slip boundary can be executed in
different ways, like shown in Fig.~\ref{fig:gradtwo}.  Here the schematic red
density gradient stays in place, while the density gradient of the other droplet
is shifted. The second gradient is drawn multiple times, transitioning from left
to right.

\begin{figure}
\includegraphics[width=\linewidth,height=0.8\linewidth]{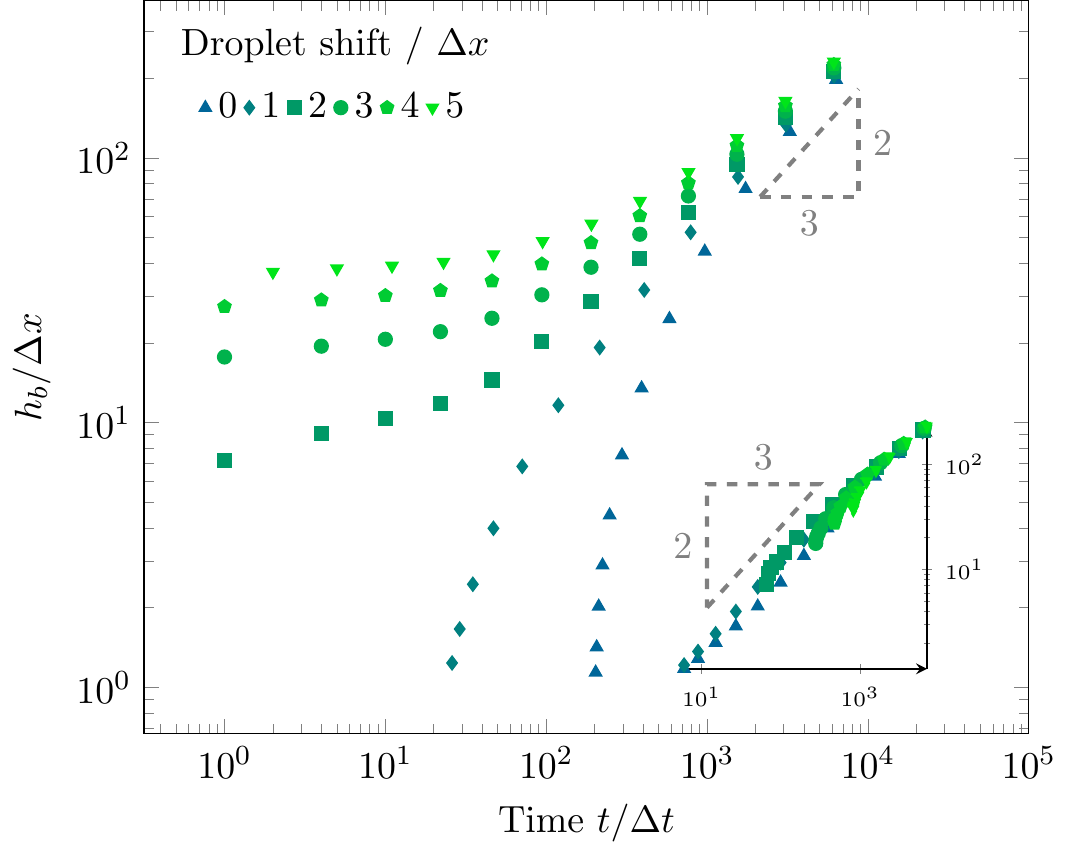}
\caption{(Color online) The meniscus height in time of differently shifted
    droplets with a contact angle of \SI{85}{\degree}.  This shift was
    implemented to trigger the coalescence after proper initialization of the
    diffuse interface.  Overlapping droplets coalesce with an initial bridge
    height $\ge0$. It can be seen that droplets with a small separation coalesce
    as well, but with a delay.  Further separation of the droplets results in a
    system where the droplets do not coalesce for $\ge10000$ timesteps. It can
    be seen that the growth rate after initial coalescence, apart from minor
    initialization effects, is identical for different droplet shifts. The inset
    displays the same data, manually shifted. It can be seen that, apart from a
    small startup phase, these curves coincide. The initial timestep of
    coalescence is therefore highly dependent on two parameters: The choice of
    the density defined as the interface position and the initial
    shift.\label{fig:tNot}}
\end{figure}

To investigate the effect of the precise location of the second drop, we
investigate the growth dynamics for different shifts, i.e.\ different initial
positions.  As an example a droplet with a contact angle of \SI{85}{\degree} is
moved by the default distance of $2$ lattice sites. The result is shown in
figure~\ref{fig:tNot}, where the bridge height $h_b$ in time $t$ for different
values of shift is recorded.  As expected, overlapping droplets coalesce with an
initial non-zero bridge height.  By contrast, small separations between the
drops cause a delay of the coalescence process by multiple timesteps. In this
case diffusion occurs across the small separation of droplets, so that it takes
some time before a detectable bridge has formed between the two drops.  Further
separation of the droplets causes the coalescence not to occur for several
thousand timesteps. After $\approx 10^3$ timesteps all curves fall on a
$t^\frac{2}{3}$ power law. In the inset of Fig.~\ref{fig:tNot} it is shown that
a collapse of the data can be achieved by manual shift of the time, which is a
robust way to identify an appropriate definition of $t=0$. This procedure will
be followed for all plots in the remainder of the paper.

\section{Results}\label{sec:results}

\begin{figure*}[tb]
   \noindent
   \begin{tabular}{@{}*{4}{p{\dimexpr .20\textwidth-.8ex}@{\hskip 
       1ex}}p{\dimexpr .20\textwidth-.8ex}}
      t=$1000 \ts$&t=$2000 \ts$&t=$3000 \ts$&t=$4000 \ts$&t=$5000 \ts$\\
      {%
          \includegraphics[height=\linewidth,width=\linewidth]{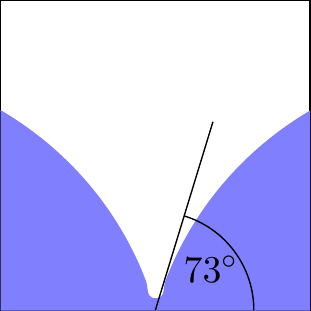}%
      }
      &
      {%
          \includegraphics[height=\linewidth,width=\linewidth]{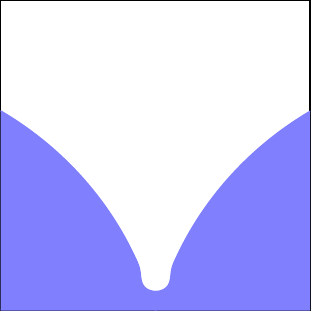}%
      }
      &
      {%
          \includegraphics[height=\linewidth,width=\linewidth]{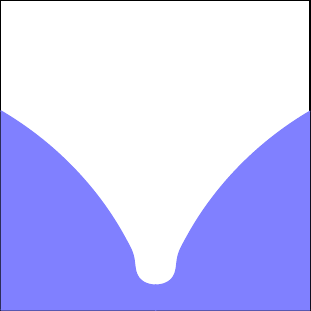}%
      }
      &
      {%
          \includegraphics[height=\linewidth,width=\linewidth]{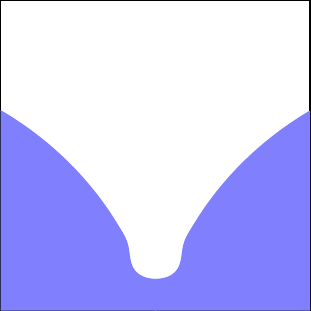}%
      }
      &
      {%
          \includegraphics[height=\linewidth,width=\linewidth]{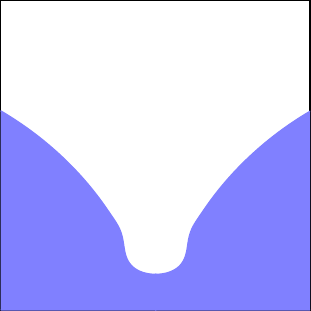}%
      }\\
      t=$2000 \ts$&t=$4000 \ts$&t=$6000 \ts$&t=$8000 \ts$&t=$10000 \ts$\\
      {%
          \includegraphics[height=\linewidth,width=\linewidth]{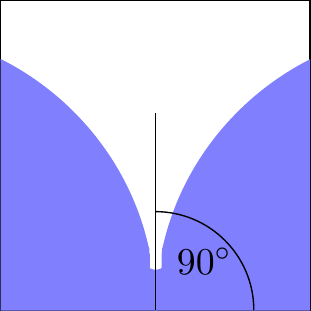}%
      }
      &
      {%
          \includegraphics[height=\linewidth,width=\linewidth]{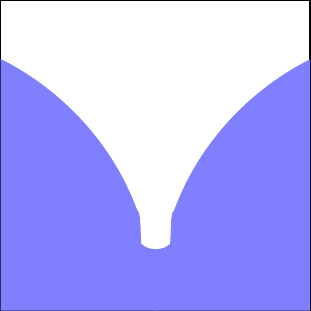}%
       }
      &
      {%
          \includegraphics[height=\linewidth,width=\linewidth]{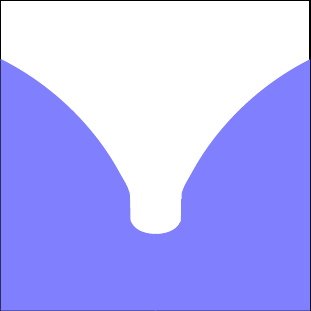}%
       }
      &
      {%
          \includegraphics[height=\linewidth,width=\linewidth]{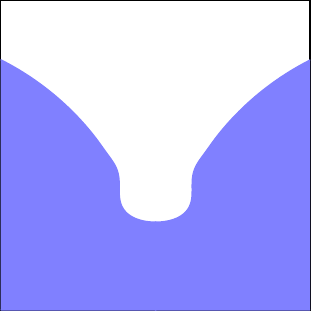}%
       }
      &
      {%
          \includegraphics[height=\linewidth,width=\linewidth]{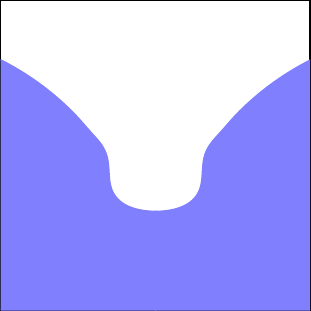}%
       }\\
    \end{tabular}

\caption{Time series of coalescence, zoomed into the meniscus shape. Upper row: 
    Droplets with a contact angle of \SI{73}{\degree} at every 1000 timesteps. Lower 
    row: Droplets with a contact angle of \SI{90}{\degree} at every 2000 timesteps. 
    The number of timesteps are displayed above the 
    snapshots.\label{fig:timeseries}}
\end{figure*}

\begin{figure}
\includegraphics[width=\linewidth]{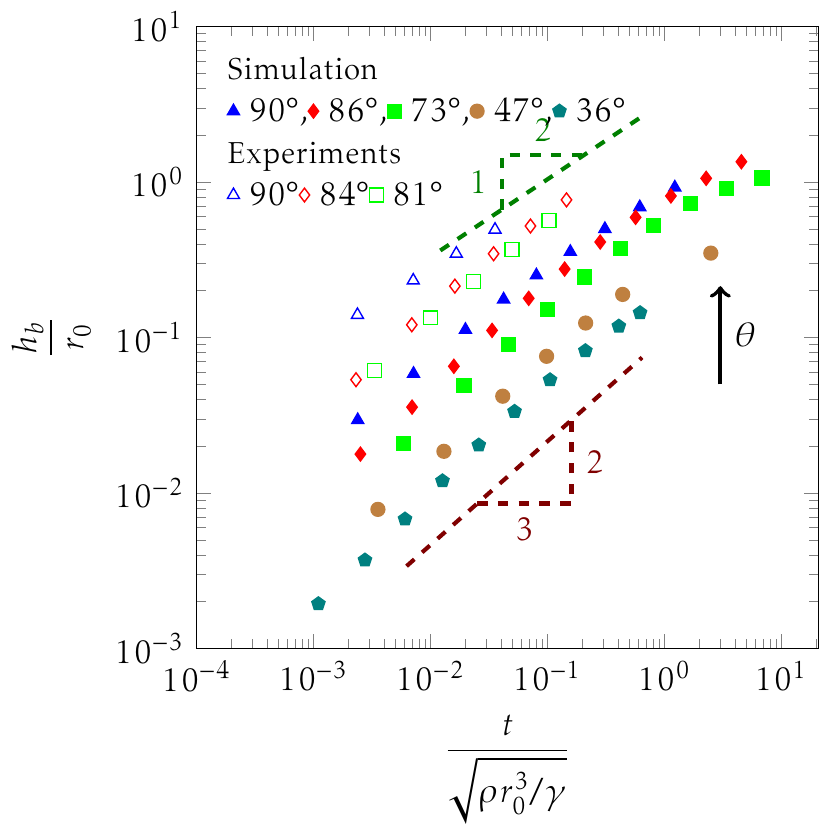}
\caption{(Color online) The bridge height as a function of time for different
    contact angles~$\theta$. The bridge height is scaled with the drop size,
    while time is rescaled by the inertio-capillary time. Closed symbols
    correspond to simulation results. Open symbols are experimental data for
    water drops in air, taken from \cite{EddWinSno13-2}.  
\new{For visual clarity, all experimental data has been multiplied with a factor of $2$, to avert overlay.}
\label{fig:scalingmatch}}
\end{figure}

In Fig.~\ref{fig:timeseries} we show snapshots of the coalescence process at the
exemplary contact angles of \SI{73}{\degree} and \SI{90}{\degree} (as is the
case in the experiments of~\cite{EddWinSno13-2}). To represent the interface we
use a bilinear interpolation of a threshold density which allows to obtain
smooth data.  The time series shows that a thin bridge appears between the two
droplets, which grows both in height and width as time evolves. To quantify this
evolution, we track the bridge height $h_b(t)$ for a broad range of contact
angles. This is shown in Fig.~\ref{fig:scalingmatch}, where we show $h_b$
(scaled with the drop radius $r_0$) as a function of  time (scaled with the
inertio-capillarity time $\sqrt{\rho r_0^3/\gamma}$). The closed symbols
represent simulations for various contact angles. For contact angles below
\SI{90}{\degree}, the initial dynamics is consistent with a $t^\frac{2}{3}$
power law until the bridge $h_b$ becomes comparable to the drop size $r_0$. This
is perfectly in line with experiments. When the contact angle is
\SI{90}{\degree}, however, the slope of the data is smaller and suggests a
smaller exponent, approaching the experimentally observed $t^\frac{1}{2}$
scaling. For a more detailed comparison, we include in
Fig.~\ref{fig:scalingmatch} the data from~\cite{EddWinSno13-2}, which
corresponds to experiments of water drops that coalesce in air. The experimental
data shown here was shifted upwards, by a factor of $2$, for the purpose of
better visualization. 
However, even without this shift the experimental data
lies about a factor $2$ above the numerical data. 
This quantitative difference
can possibly be attributed to the fact that the simulations consider droplets
that are immersed into an outer fluid of equal density. The transport inside the
outer fluid does slow down the dynamics with respect to the case of drops in
air, which is consistent with the observations in  Fig.~\ref{fig:scalingmatch}.
    Therefore a quantitative match of these is not to be expected, but
    in terms of scaling laws the simulations comply with experiments. 
    As the wettability of the substrate alters the contact angle of the fluid,
    it alters the capillary pressure of the droplets.
    Due to the capillary pressure driving the coalescence,
    the contact angle alters the rate at which the meniscus grows.
    The representative scaling laws for this behavior, as well as the behavior of the fluid interface are discussed below.

\subsection{Results for $\theta < \SI{90}{\degree}$}

To make further use of the simulation results, let us briefly revisit the usual
scaling arguments for coalescence. The situation is best understood for contact
angles $\theta < \SI{90}{\degree}$, for which the horizontal scale and vertical
scale are simply proportional to one another: the ratio of the two lengths is
set by the tangent of the contact angle. This can for example be seen from
Fig.~\ref{fig:timeseries}, showing that the width of the meniscus increases as
well as $h_b$ during the growth. Since $h_b$ sets the characteristic scale of
the bridge, the capillary pressure can be estimated as 

\begin{equation}
   P_{\text{cap}}\propto\frac{\gamma}{h_b}.
\end{equation}
Similarly, the inertial pressure is obtained as

\begin{equation}
   P_{\text{iner}}\propto\rho\left(\frac{h_b}{t}\right)^2,
\end{equation}
which then leads to the observed
\begin{equation}
   h_b\propto t^\frac{2}{3}.
\end{equation}

\begin{figure}
\includegraphics[width=\linewidth]{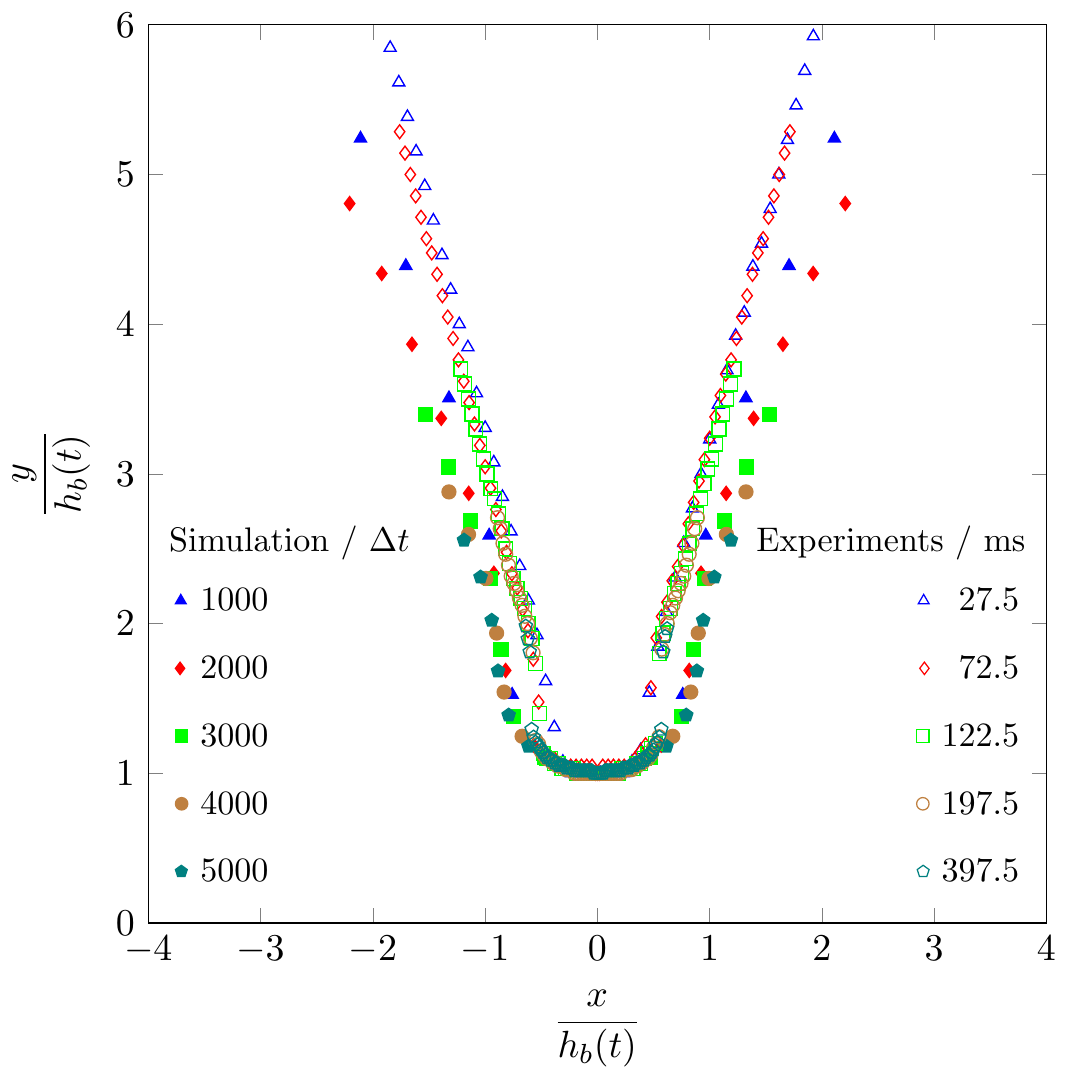}
\caption{(Color online) \label{fig:SelfSim73}Rescaled bridge shape for droplets 
with a contact angle of \SI{73}{\degree} reveal a self-similar bridge growth. 
Closed symbols correspond to simulation results, 
open symbols are experimental data for water drops in air \cite{EddWinSno13-2}.}
\end{figure}

To further test the idea that the dynamics is governed by the growing length
scale $h_b(t)$, one can attempt a collapse of the bridge profiles during the
growth process. This is shown for the case $\theta=\SI{73}{\degree}$ in
Fig.~\ref{fig:SelfSim73}, where we overlay the meniscus shapes for different
times, after rescaling the horizontal and vertical scales with $h_b(t)$. The
scaled profiles indeed exhibit an excellent collapse. This confirms that the
bridge growth is characterized by a universal spatial profile, and that the
temporal dependence can be effectively absorbed in the growing length scale
$h_b(t)$. The self-similarity only applies for the initial stages of
coalescence, so the data shown are until the bridge height reaches about one
third of the initial drop height. 
    This limits the data to parts of the droplet deformed by the coalescence,
    where the scaling law is applicable.
    Small deviations far from the meniscus can be attributed to this effect.
Figure~\ref{fig:SelfSim73} also shows the
corresponding experimental plot from~\cite{EddWinSno13-2}, for which
self-similarity was convincingly demonstrated as well. The numerical bridge
shapes (for immersed drops) differ slightly from the experimental profiles, but
the same principle of self-similarity is valid during the initial stages of
coalescence. 

\begin{figure}
\includegraphics[width=\linewidth]{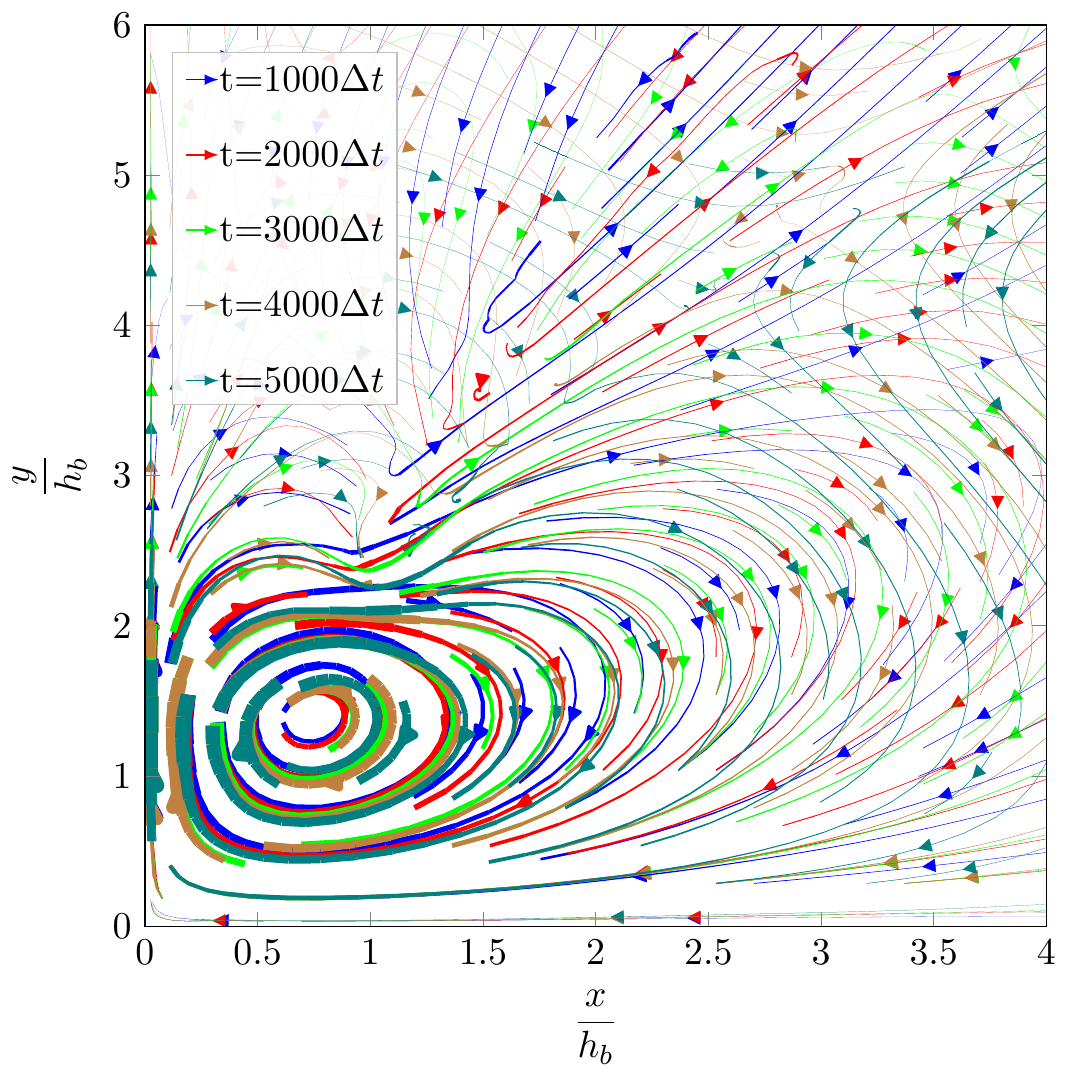}
\caption{(Color online) Streamlines of the velocities in coalescing droplets 
    with a contact angle of \SI{73}{\degree}, to the right of the symmetry axis. 
    The positions of the velocities are rescaled like the interface positions. 
    The amplitude of the velocities is scaled as $v_x h_b^{\frac{1}{2}}$ and 
    $v_y h_b^{\frac{1}{2}}$. The amplitude of the field is shown by varying 
    the width of the streamlines linearly with the amplitude of the underlying vectors.
\label{fig:vel73}}
\end{figure}

Interestingly, the simulations allow one to extract information that is not
easily accessible through experiments, such as the fluid velocities $v_x$ and
$v_y$. Inspired by the idea of self-similarity, we rescale the streamline
patterns by again normalizing $x$ and $y$ by $h_b(t)$. The result is shown in
Fig.~\ref{fig:vel73}, where the streamlines are obtained after scaling the
velocities with $h_b^{\frac{1}{2}}$ (different times are visualized using
different colors). The streamline patterns exhibit an excellent collapse. It can
be seen that the coalescence causes a recirculating flow, with a vortex located
to the right of the symmetry axis, driving the meniscus upwards. The necessary
rescaling of the velocity vectors can be understood by considering $u \propto
h_b/t \propto t^{-1/3}$. This implies that $u h_b^{1/2} = {\rm const.}$, as was
indeed used in preparing Fig.~\ref{fig:vel73}.

\subsection{Results for $\theta = \SI{90}{\degree}$}

Let us now turn to the case of droplets with $\theta= \SI{90}{\degree}$, for
which the interfaces are tangent when brought into contact. As a consequence of
this geometry, the horizontal and vertical scales are no longer the same. We
therefore introduce the width of the bridge $w$, indicated in
Fig.~\ref{fig:Experiment}, as the horizontal scale that is much smaller than
$h_b$. The geometry is such that 

\begin{equation}
   w\propto \frac{h_b^2}{r_0},
\end{equation}
and subsequently, the scaling laws need to account for this disparity of
horizontal and vertical scales. The usual argument is that the capillary
pressure reads

\begin{equation}\label{eq:pcap}
   P_{\text{cap}}\propto\frac{\gamma}{w} \propto\frac{\gamma r_0}{h_b^2},
\end{equation}
which can be balanced with the inertial pressure

\begin{equation}\label{eq:piner}
   P_{\text{iner}}\propto\rho\left(\frac{h_b}{t}\right)^2.
\end{equation}
This leads to

\begin{equation}
   h_b\propto t^\frac{1}{2},
\end{equation}
and explains why the meniscus growth differs from the 2/3 law observed for
smaller contact angles (cf. Fig.~ \ref{fig:scalingmatch}).

\begin{figure}
\includegraphics[width=\linewidth]{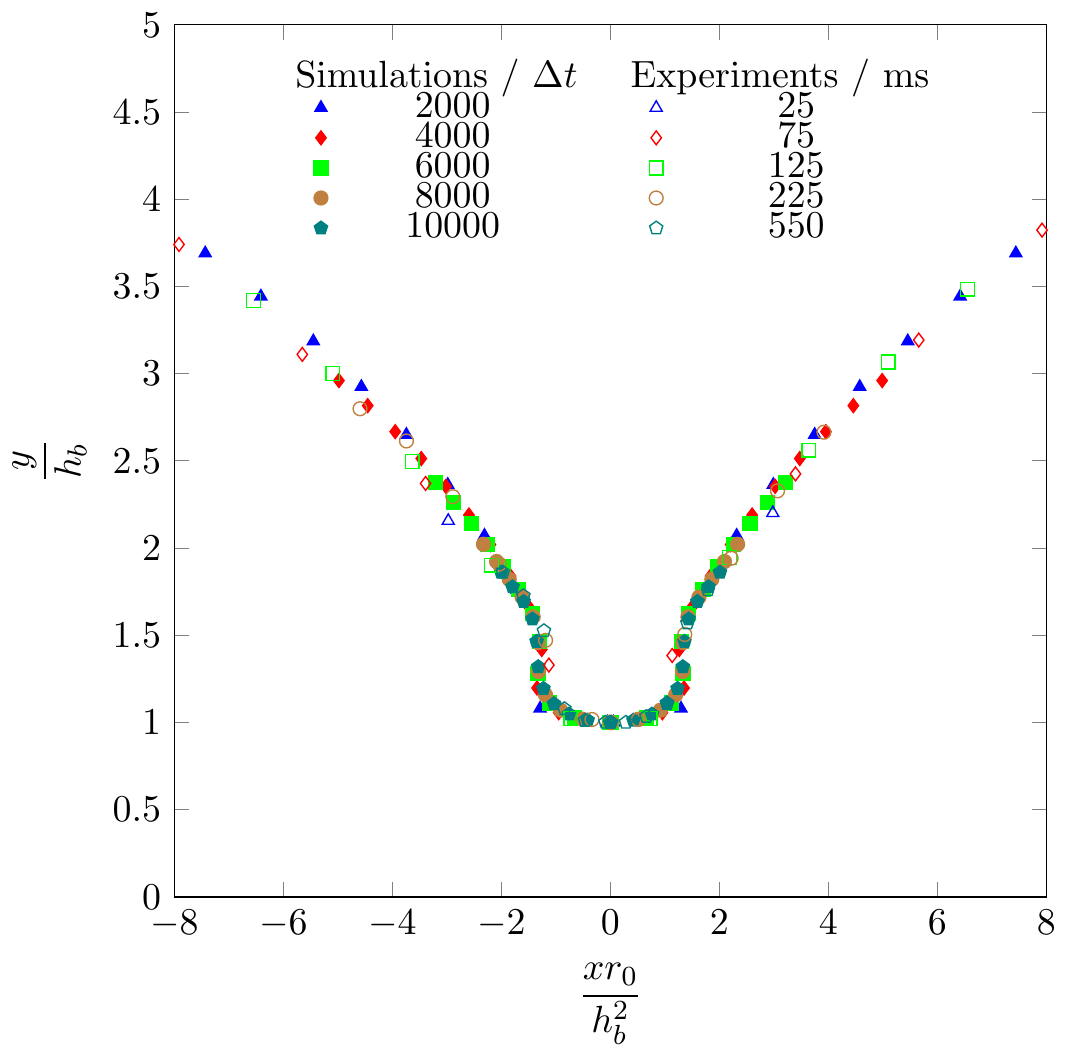}
\caption{(Color online) Rescaled bridge shape for droplets 
with a contact angle of \SI{90}{\degree} reveal a self-similar bridge growth. 
Note that the horizontal and vertical axis are scaled differently. 
Closed symbols correspond to simulation results, 
open symbols are experimental data for water drops in air \cite{EddWinSno13-2}.}
\label{fig:SelfSim90}
\end{figure}

\begin{figure}
\includegraphics[width=\linewidth]{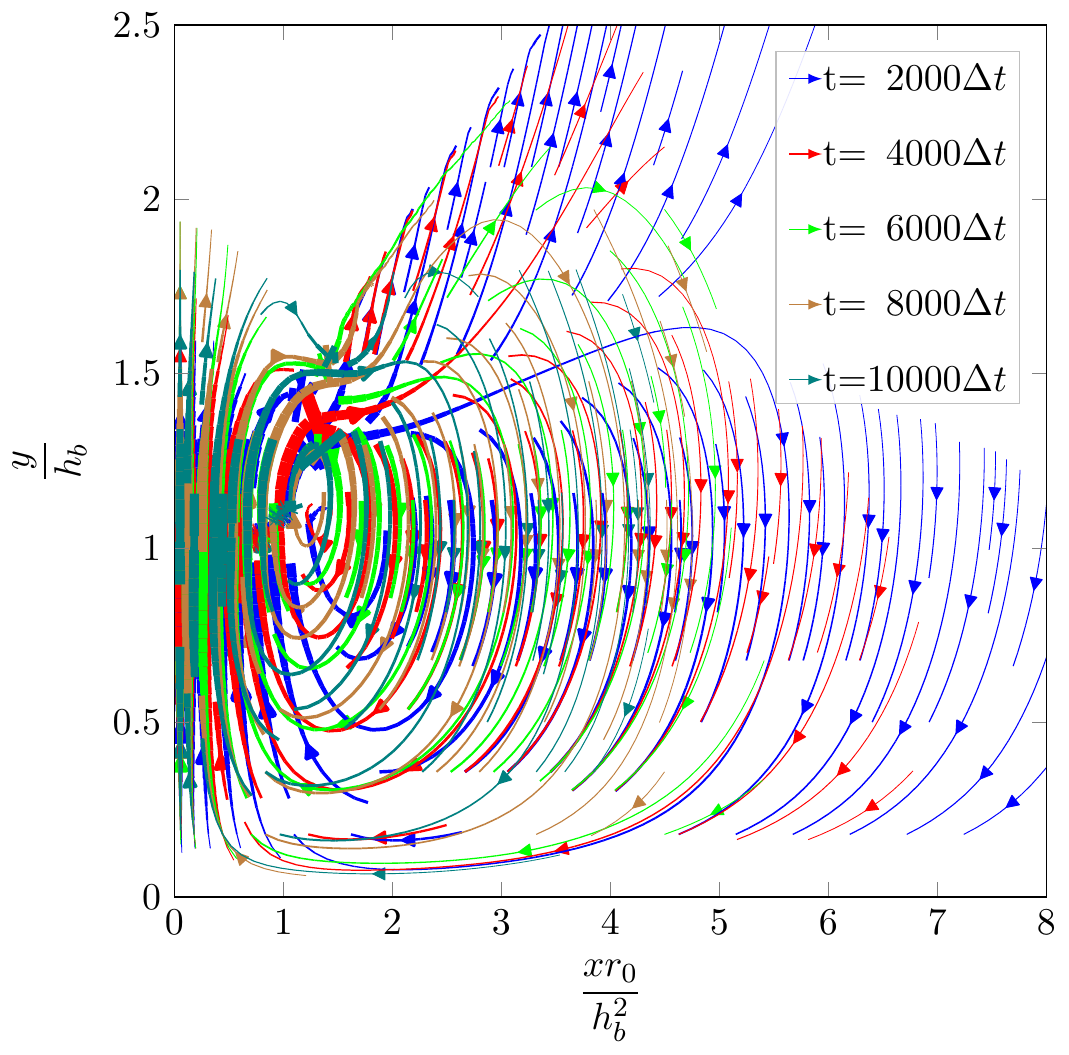}
\caption{(Color online) Streamlines of the velocities in coalescing droplets 
    with a contact angle of \SI{73}{\degree}, to the right of the symmetry axis.
    The positions of the velocities are rescaled like the interface positions.
    The  amplitude of the velocities is scaled with the empirical values of $v_x
    h_b^{\frac{1}{3}}$ and $v_y h_b^{\frac{1}{2}}$. The amplitude of the field
    is shown by varying the width of the streamlines linearly with the amplitude
    of the underlying vectors.  \label{fig:vel90}}
\end{figure}

Once more, we will test these scaling ideas by searching for self-similar
dynamics, both for the bridge shape and for the velocity profiles. The first of
these tests is provided in Fig.~\ref{fig:SelfSim90}, where the bridge profiles
for $\theta= \SI{90}{\degree}$ are rescaled with $w \sim h_b^2/r_0$ on the
horizontal axis and with $h_b$ in the vertical direction. A collapse is indeed
observed, confirming the necessity of taking different horizontal and vertical
scales. In addition the numerical profiles exhibit a perfect agreement with the
experimental results for the bridge shape~\cite{EddWinSno13-2}.
Intriguingly,
however, we have not been able to obtain a convincing self-similarity for the
velocity fields for $\theta= \SI{90}{\degree}$. Following the logic above, one
would expect for the horizontal velocity $v_x \propto w/t \propto h_b^2/R t
\propto t^0$, while for the vertical velocity $v_y \propto h_b/t \propto
t^{-1/2}$. However, the best ``collapse" was obtained by empirically scaling the
velocities respectively as $v_x h_b^{\frac{1}{3}}$ and $v_y h_b^{\frac{1}{2}}$,
and the result is shown in Fig.~\ref{fig:vel90}. One again observes a
recirculating flow that leads to the bridge growth, but the associated vortex
structure is not perfectly self-similar. In particular, we note that the vortex
appears to become smaller in time, after rescaling, suggesting that $h_b$ and
$w$ are not the correct scales for the velocity field.  The velocities of
strongest amplitude lie underneath the meniscus and are nearly only vertical.
Mass conservation in this case is achieved by enlarging the respective area in
time.  This means that the usual scaling arguments of equations \ref{eq:pcap}
and \ref{eq:piner} might actually be too simplistic. For example, the
self-similar scaling of the meniscus profiles implies that the typical curvature
scales as $h_b/w^2 \sim r_0^2/h_b^3$, and not as $1/w \sim r_0/h_b^2$, as was
assumed in  (\ref{eq:pcap}). This would change the coalescence exponent from 1/2
to 2/5, which does not concur with simulations and previous experiments.
    This observation on the velocity field suggests that the pressure scaling (\ref{eq:piner}) needs to be revised.
    The flow field is more intricate than the scaling argument allows to believe.
    Uncertainties in the definition of the scaling argument undermine this requirement.
    This would be an interesting topic for future work, for which a larger range of numerical data would be required to conclusively infer the relevant scaling laws. 
    Droplets with a contact angle of \SI{90}{\degree} only differ from freely suspended coalescing droplets by a minor amount of surface friction.
    Therefore the scaling argument for freely floating droplets might need to be revisited as well.

\section{Conclusion}\label{sec:discussion}
We simulated the coalescence of submerged droplets with different
contact angles and compared our results to experimental data.  Similar growth
rates for the bridge height in time, equally dependent on the contact angle
could be found in this case. Despite quantitative differences of the interface position
of the fluid interface between experimental data of droplets in air and the
simulation of submerged droplets, the same rescaling argument revealed a self
similarity in time. 
Being able to use the same scaling arguments to scale the interface position of both the experimental droplets in air and the
submerged simulated droplets shows the universality of the scaling argument.
We applied this scaling law to the velocity field and, for droplets of a \SI{73}{\degree} contact angle, revealed the underlying velocities that cause the coalescence and give reasons for the scaling of the amplitude of the velocities.  
For droplets of a \SI{90}{\degree} contact angle we presented that the velocities causing the coalescence are more intricate than the scaling laws indicate.
This shows that, though these scaling laws seem to work for the interface position, the underlying estimate for the relevant scaling for the velocity appears to be inconsistent with the internal flow structure. Clearly, our simulations show that the droplet internals are more complex than usually assumed.
\new{Our findings have implications for the design of devices in open microfluidics where different fluids are transported on chemically patterned substrates. Understanding the formation, transport and the coaleascence of droplets in particular is mandatory to optimize these devices and ascertain a reliable long-term functionality.}

\section*{Acknowledgment}
This research was carried out under project number M61.2.12454b in the framework 
of the Research Program of the Materials innovation institute M2i (www.m2i.nl).
We highly acknowledge Oc\'e-Technologies B.V.\ for financial support and the
J\"ulich Supercomputing Centre for the required computing time.


\end{document}